\documentclass{aastex631}

\usepackage{forest}
\usepackage{amsmath}
\usetikzlibrary{fit,positioning}
\tikzset{
  red arrow/.style={
    midway,red,sloped,fill, minimum height=3cm, single arrow, single arrow head extend=.5cm, single arrow head indent=.25cm,xscale=0.3,yscale=0.15,
    allow upside down
  },
  black arrow/.style 2 args={-stealth, shorten >=#1, shorten <=#2},
  black arrow/.default={1mm}{1mm},
  tree box/.style={draw, rounded corners, inner sep=0.5em},
  node box/.style={white, draw=black, text=black, rectangle, rounded corners},
}

\shorttitle{Predicting the energetic proton flux with machine learning}
\shortauthors{Stumpo et al.}

\graphicspath{{./}{figures/}}

\begin{document}

\title{Predicting the energetic proton flux with a machine learning regression algorithm}

\author[0000-0002-6303-5329]{Mirko Stumpo}
\affiliation{INAF-Istituto di Astrofisica e Planetologia Spaziali, 00133 Roma, Italy}

\author[0000-0001-5481-4534]{Monica Laurenza}
\affiliation{INAF-Istituto di Astrofisica e Planetologia Spaziali, 00133 Roma, Italy}

\author[0000-0002-7102-5032]{Simone Benella}
\affiliation{INAF-Istituto di Astrofisica e Planetologia Spaziali, 00133 Roma, Italy}

\author[0000-0002-5002-6060]{Maria Federica Marcucci}
\affiliation{INAF-Istituto di Astrofisica e Planetologia Spaziali, 00133 Roma, Italy}

\begin{abstract}

The need of real-time of monitoring and alerting systems for Space Weather hazards has grown significantly in the last two decades. One of the most important challenge for space mission operations and planning is the prediction of solar proton events (SPEs). In this context, artificial intelligence and machine learning techniques have opened a new frontier, providing a new paradigm for statistical forecasting algorithms. 
The great majority of these models aim to predict the occurrence of a SPE, i.e., they are based on the classification approach. In this work we present a simple and efficient machine learning regression algorithm which is able to forecast the energetic proton flux up to 1 hour ahead by exploiting features derived from the electron flux only. This approach could be helpful to improve monitoring systems of the radiation risk in both deep space and near-Earth environments. The model is very relevant for mission operations and planning, especially when flare characteristics and source location are not available in real time, as at Mars distance.
\end{abstract}


\section{Introduction} \label{sec:intro}
Solar Proton Events (SPEs) are pronounced enhancements of the energetic proton flux measured by instruments placed on different space probes across the Heliosphere. Solar protons can reach high energies, say tens of GeVs, as a consequence of different 
acceleration processes occurring at the Sun 
in association with transient phenomena like solar flares and coronal mass ejections \citep[CMEs;][]{kahler1984association,shea1990a,Aschwanden2002,iucci2005satellite}. Then, particles travel along interplanetary magnetic field lines and can produce a geoeffective SPE that can be detected by instruments placed on Earth-orbiting satellites, such as the Geostationary Operational Environmental Satellite (GOES). Geoeffective SPEs are defined operationally as $\geq$S1 events, i.e., when the $>$10 MeV proton flux overcomes the threshold of 10 pfu. When SPEs reach relativistic energies, secondary particles are produced in the Earth atmosphere, causing sharp increases in the count rates of ground-based particle detectors, called ground level enhancements \citep{forbush1946three}. SPEs have been traditionally divided in impulsive and gradual, depending on their time profile after the onset and on different properties such as composition and acceleration mechanisms. Nevertheless, a more complex scenario emerged after the launch of Advanced Composition Explorer (ACE) and Solar and Heliospheric Observatory (SOHO) in the late 90s, when hybrid events were identified, possessing properties of the both fundamental types, leading to an expanded classification system \citep[e.g.,][]{Cohen1999, Kocharov2002, Cane2003, Cliver2008}. 
The primary acceleration source producing protons with energies $>$10 MeV is still a subject of debate \citep[e.g.,][and references therein]{zhang2020earth}, with theories suggesting CME-driven shocks or post-CME reconnection processes \citep{Cane2003,reames2000particle,desai2016large}.

From a Space Weather perspective, SPEs pose significant hazards within the interplanetary and near-Earth space \citep{eastwood2017economic}. Indeed, they can damage spacecraft instruments \citep{iucci2005space}, interfere with radio communications in Earth's atmosphere and spacecraft operations \citep{iucci2005satellite}. Moreover, they present radiation threats to astronauts \citep{furukawa2020space}, and even to airline crews, and passengers that traverse polar routes when they reach relativistic energies \citep{jones2005space}. Therefore, the development of warning systems is vital to predict the occurrence of SPEs and ensure an effective mitigation against them. Such systems would involve real-time monitoring of the solar activity, and/or advanced modeling of SPE generation and transport, as well as implementing dedicated safety protocols for the protection of technologies and human health. The need of forecasting systems or at least monitoring of the radiation environment is increasingly growing also in view of the new era of sustainable human space exploration \citep[e.g. Moon and Mars;][]{zhang2023radiation, fogtman2023towards}. Indeed, no natural shielding like the Earth magnetosphere is present in the deep space. Radiation exposure may cause several diseases to humans \citep{furukawa2020space}, including the induction of cataract \citep{cucinotta2001space}, cancer \citep{sridharan2015understanding,cucinotta2015review,chang2016harderian}, damages to nervous and heart systems \citep{jandial2018space, hughson2018heart} and acute radiation syndrome \citep{wu2009risk,carnell2021spaceflight}. Recent studies have identified that space radiations also affect plants, which are an important part of the environment that would support humans for the long-term habitation of spacecraft \citep{ferl2002plants, arena2014space}.
Therefore, the planning of this type of missions must include a radiation risk assessment, including the warning system for SPEs \citep{fogtman2023towards}.

The prediction of SPEs constitutes a challenging task due to different factors at the origin of the particle acceleration, including solar conditions, seed particles, shock geometry, coronal field configuration, and interplanetary transport. Moreover, a main hurdle for SPE forecasting is the short transit time of protons travelling from Sun to Earth, going from $\sim$15 min up to few hours in the energy range between 10 MeV and 20 GeV. Several models devoted to the forecasting of SPEs have been introduced in the last decades, as reported in a comprehensive review by \cite{Whitman2023}. SPE forecasting models can be mainly divided into two classes: \textit{physics-based} models, and \textit{empirical} models. In recent years, the latter class of models has also taken advantage from the application of machine learning algorithms, thanks to which it is possible to recognize hidden characteristics in the data and use them to automatically adapt the prediction to changing physical conditions. 

Physics-based models  usually simulate processes of particle release/acceleration at the source and transport/diffusion in the interplanetary space.
A significant shortcoming of these models is the high computational cost needed to achieve comprehensive and accurate numerical simulations, thus 
making them operationally challenging. In addition, their performances strongly depend on the precision of the input data and on the validity of the underlying physical assumptions. In recent years, significant efforts have been made with the goal of improving the performance of physics-based models and making them suitable for operational forecasting. This is the case of the SEPMOD for instance, which approximates the observed SPEs through a simple shock source description and a scatter-free particle propagation \citep{Luhmann2007,Luhmann2010}, later integrated with the Wang–Sheeley–Arge coronal model \citep{Arge2004} and with the 3D Enlil heliospheric model \citep{Odstrcil2003}, allowing for simulating the ICME propagation \citep{Palmerio2022}. 

Empirical models 
primarily rely on identifying statistical relations between SPEs and possible precursors. These models leverage historical data and statistical analysis to make predictions, instead of performing numerical simulations grounded on the knowledge of the physical processes involved. 
Hence, models belonging to this class are computationally less demanding and more {suitable} in operational settings.
Early models to predict SPEs include: the ``proton prediction system" \citep[PPS76;][]{Smart1979,smart1989pps,kahler2007validating} based on solar flare parameters (microwave/X-ray flux, flare location, etc.); the Protons model \citep[][currently in use on the NOAA Space Weather Prediction Center]{Balch2008} using also the occurrence of radio type II and type IV bursts as input parameters, being indicative of the presence of a CME driven shock.
The Empirical model for Solar Proton Events Real Time Alert \citep[ESPERTA;][]{laurenza2009technique,alberti2017solar} performs a logistic regression analysis on three input parameters, namely, the flare location, the 1–8 \AA soft X-ray (SXR) fluence, that is indicative of flare strength, and the 1 MHz radio fluence, which allows for the identification of Type III radio bursts, indicative of fast electron beams escaping from the acceleration site into the interplanetary space. ESPERTA aims to provide timely warnings within a time interval of 10 minute following the SXR peak time of flares of $\geq$M2 class. It has also been adapted to forecast only the largest radiation storms ($\ge$S2) produced by $\geq$100 pfu SPEs \citep{laurenza2018short}, and reinterpreted within a machine learning framework \citep{Stumpo2021,Laurenza2024}.
Other empirical models incorporate also information about CMEs 
\citep[e.g.,][]{papaioannou2016solar,st2017solar,anastasiadis2017predicting,papaioannou2018nowcasting,papaioannou2022the} or exploit the arrival times of relativistic electrons and/or high-energy protons at 1 AU \citep[e.g., REleASE and UMASEP;][]{posner2007up,Nunez2011predicting}. 
The REleASE model initially relied only on SOHO data, and it has been adapted later to use also ACE observations \citep{Malandraki2018}. The UMASEP scheme, introduced by \cite{Nunez2011predicting,nunez2015real}, is based on the lag-correlation between the time derivatives of the SXR flux and the near-Earth proton differential fluxes (e.g., those provided by GOES). 
Machine learning models, such as decision trees, random forest and neural networks, have been introduced in the landscape of SPE forecasting models \citep{boubrahimi2017prediction,engell2017sprints,nunez2020predicting, lavasa2021assessing, hosseinzadeh2024improving, rotti2024short}.


The models introduced so far aim to predict mainly whether a SPE will occur or not and/or some SPE properties.
Hence, generally they do not provide the forecast of the proton flux, which is a slightly different problem in terms of model development and search for input features. Several authors privileged the flux prediction, instead of making classification, providing a timely estimate of the proton flux and also providing the possibility of more detailed analysis in an operational setting. Such methodologies have recently been also introduced in the time series reconstruction of cosmic-ray intensity in the heliosphere \citep[e.g., see][]{Sabbatini2022,Sabbatini2024}. Concerning the SPE forecasting, recently \cite{nedal2023forecasting} employed the bidirectional long short-term memory (bi-LSTM) model architecture in order to make predictions of the daily $>10$Mev, $>30$MeV and $>60$MeV proton flux. They take as input features the sunspot number, the F10.7 index, the X-ray flux, the solar wind speed, and the interplanetary magnetic field magnitude. \cite{Laurenza2024} introduced an LSTM-based algorithm for predicting the GOES $>10$ MeV proton flux 1 hour in advance based on measurements of the electron flux data in the energy range 0.25–0.7 MeV obtained from the Electron Proton Helium INstrument (EPHIN) on board the Solar and Heliospheric Observatory (SOHO). Here, authors compare the scores of the LSTM-based model with the ESPERTA classifier, showing that the former significantly improves the performance, lowering the false alarm rate, albeit the reduced warning time due to the introduction of the electron flux as main input feature.

The great majority of the existing models reviewed so far are based on flare and cme precursors/characteristics.
However, warning systems 
for supporting space exploration missions are even more challenging because flare precursors for predicting SPEs cannot be always exploited. For example at Mars distance, the nominal magnetic field line connection to the Sun is at $\sim$ W90, implying that the full flare signature may not be visible as the source region could be often located behind the solar limb \citep{posner2020warning}.
In addition, communications delays or gaps require  
innovative and high-performance methods entirely based on \textit{in situ} data for future space missions designed for Space Weather activities and beyond. With this aim, we introduce a novel machine learning model based on the random forest (RF) algorithm for energetic proton flux forecasting. The main goal of this work is to create a model capable of giving an estimation of the energetic proton flux in a desired energy range based on near relativistic electron flux. The model production pipeline can be adapted to build and train similar models involving integral proton flux (e.g. $>$10 MeV) or differential proton flux. This makes our approach very helpful in view of space missions carrying on board particle detectors that allows for electron and proton flux measurements, with the option of providing a dedicated SPEs prediction system to be directly implemented on board.
Therefore, this work contribution may be twofold. Firstly it contributes to the need of efficient and accurate algorithms for predicting the proton flux forward in time directly onboard. Secondly, this work may be helpful for designing warning systems not based on flare characteristics, thus potentially supporting space exploration missions.

\section{Dataset}
In real-time monitoring and space weather related contexts, several space missions collected a huge amount of data during the last solar cycles which are sufficient for training machine learning algorithms. 
In this work we used data from the COSTEP/EPHIN instrument placed on board the ESA/NASA SOHO spaceprobe, orbiting around the L1 libration point \citep[see][for a description of the COSTEP experiment]{Muller-Mellin1995}. The EPHIN telescope is meant for measuring electrons in the energy range from 250 keV to $>$8.7 MeV, and protons and alpha particles from 4 MeV/nucleon to $>$53 MeV/nucleon (see Table \ref{Tab:tab1}) with a 1 minute cadence. Note that SPEs definition is usually based on GOES integral proton flux ($>10$ MeV), thus our results are not directly comparable with the intensities reported by GOES. However, in order to show the capabilities of our approach, we integrated the COSTEP/EPHIN differential channels and merged the P8, P25 and P41 channels. We compared the resulting time series with the GOES $>10$ MeV integral channel and they are very similar in terms of both flux value and profile (not shown).
Thus the energetic proton flux we are going to consider in this work is in the range of energies between 7.8 MeV and 53 MeV. The COSTEP/EPHIN instrument is also provided of an integral channel measuring together i) $>53$ MeV protons and alpha particles; ii) $>8.7$ MeV electrons. We did not include this channel in the merged one because of the presence of both electrons and alpha particles together with protons.

\begin{table}
    \centering
    \begin{tabular}{c|c|c} 
        Channel & Species & Energy range \\
        \hline
        E150 & Electron & (0.25 - 0.70) MeV \\
        E300 & Electron & (0.70 - 3.00) MeV \\
        E1300 & Electron & (2.60 - 6.20) MeV \\
        E3000 & Electron & (4.80 - 10.40) MeV \\ 
        P4  & Proton & (4.30 - 7.80) MeV \\
        P8  & Proton & (7.80 - 25.00) MeV  \\
        P25 & Proton & (25.00 - 40.90) MeV \\ 
        P41 & Proton & (40.90 - 53.00) MeV\\
        \text{INT} & p + e + He & $>$53 MeV \\
        \hline
    \end{tabular}
    \caption{Proton/electron differential channels and the integral channel (INT) of the EPHIN instrument. Adapted from \citet{muller1995costep}.}
    \label{Tab:tab1}
\end{table}

\begin{figure}[!h]
   	\centering
    \includegraphics[width=\columnwidth]{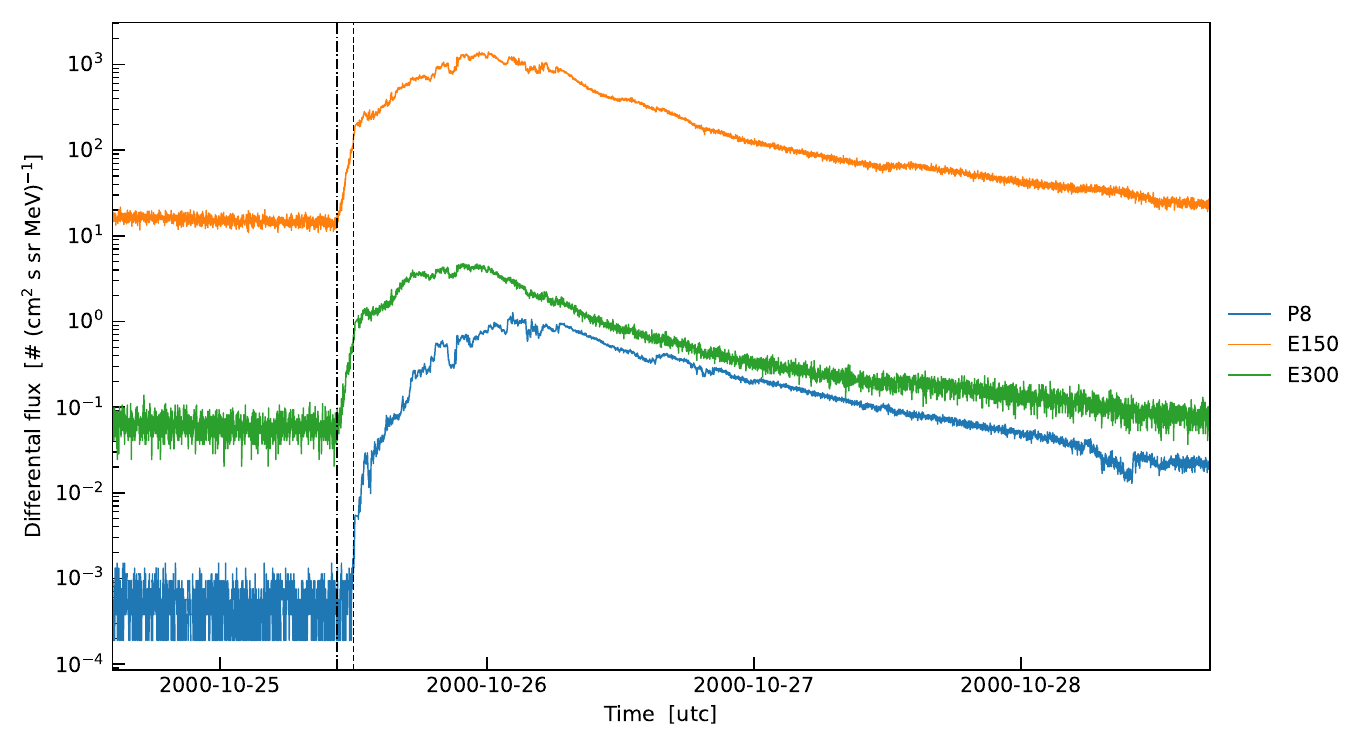}
    \caption{EPHIN proton/electron differential channels P8 (blue), E150 (orange) and E300 (green) during the event of October 25, 2000. The black vertical lines mark the onset in proton and electron data. In this event, the electrons measured in E150 and E300 differential channels arrive 01:30 hours earlier than protons in the P8 channel. Furthermore, the time profiles are very similar.}
   	\label{fig:data}
\end{figure}

In order to make reliable and efficient predictions for increases in the energetic proton flux due to solar particles we need information about the magnetic connection between the source and the observer. Some classification algorithms, based on flare precursors \citep[e.g.,][]{laurenza2009technique,Stumpo2021}, use the heliolongitude of the flare, inferred from observations of the solar flare region in the $H\alpha$ spectral line. In this work the information on magnetic connectivity is indirectly provided by the earlier arrival of quasi-relativistic electrons. Therefore, we base our model on the hypothesis that electron and proton events share the same source region. Figure \ref{fig:data} shows an event included in the dataset used in this work. The onset of the event in the near-relativistic electron data anticipates the one in the proton data at several MeVs, as expected. Furthermore, the time profiles of electrons and protons are very similar \citep[see also][]{posner2007up}.

\subsection{Feature selection}\label{sec:feature_selection}

The target of the model is the logarithm of the proton flux integrated in the channels P8, P25 and P41 of the COSTEP/EPHIN instrument (see Table \ref{Tab:tab1}). From now on, for the sake of simplicity, we will refer to logarithm of the integrated proton flux at such energies as only proton flux. The physical observables used in the training set are:
\begin{itemize}
    \item electron differential channel E150 (0.25-0.70) MeV
    \item electron differential channel E300 (0.67-3.00) MeV
    \item electron differential channel E1300 (2.64-6.18) MeV
    \item electron differential channel E3000 (4.80-10.4) MeV
    \item derivative of each electron differential channel
    \item derivative of the proton flux
\end{itemize}
All these quantities have been log-transformed in order to reduce the range of variability of data between the time profiles produced by background protons and SPEs. 
The derivatives of i) each electron differential channel and ii) the proton flux, are here considered in the training dataset in order to provide the model of the information about the rise parameter/grow rate \citep[and thus about the magnetic connectivity, indirectly][]{posner2007up}. Note that the log-transformation has been applied to the proton and electron fluxes before calculating the derivatives in order to avoid negative values.
Each feature in principle may appear with different temporal lags with respect to the proton flux (as also explained in Section \ref{sec:rf}). 
Finally, the training dataset is completed by the average and the standard deviation of each physical observable, computed in moving windows of different widths. In particular, we set the number of windows to be the same as the number of lags $n_{\text{lags}}$ and their temporal widths are the same as the lags. 
Therefore, the total number of features is given by $N = 27 \cdot n_{\text{lags}}$, where $27=9+9\cdot 2$, i.e., obtained by summing up the 9 physical observables and the 9   physical observables multiplied by the 2 statistic measures (mean and standard deviation) considered for each feature.
However, even though the procedure described above is quite general, we choose to use only one lag (1 hour) in order to avoid overfitting. Thus we have a total of $N=27$ features and the lookback is $t_b = 1$ hour. Regularization techniques are necessary for longer lookbacks, which are not currently implemented in our pipeline.

In principle any lookforward can be chosen. Our choice ($t_f=1$ hour) is motivated by the onset delay analysis reported by \citet{posner2007up}, according to which the forecasting time of the energetic proton flux based only on the relativistic electron flux is up to 1 hour. In general, there is a trade-off between the lookback and the lookforward of the model. The prediction becomes smoother and more accurate as the number of lags of the features increases, but the model tends to overfit the training dataset. For example a time offset between the target and the prediction may appear. 

\begin{table}
    \centering
        \begin{tabular}{c| c}
        \hline
        \hline
        $\mathbf{n_{tree}}$ & 50\\
        \hline
        \textbf{max. depth} & 50\\
        \hline
        \textbf{criterion} & Squared error\\
        \hline
        \textbf{max features} & sqrt\\
        \hline
        \textbf{min. samples split} & 10 \\
        \hline
        \textbf{min. samples leaf} & 10\\
        \hline
        \hline
    \end{tabular}
    \caption{Hyperparameters of the RF regressor used in this work, referred to the Python implementation given by scikit-learn \citep[version-1.2.2;][]{scikit-learn}. All the non-mentioned hyperparameters are kept to the default values.}
    \label{tab:tab_rf_pars}
\end{table}

\section{Random Forest Regressor}\label{sec:rf}

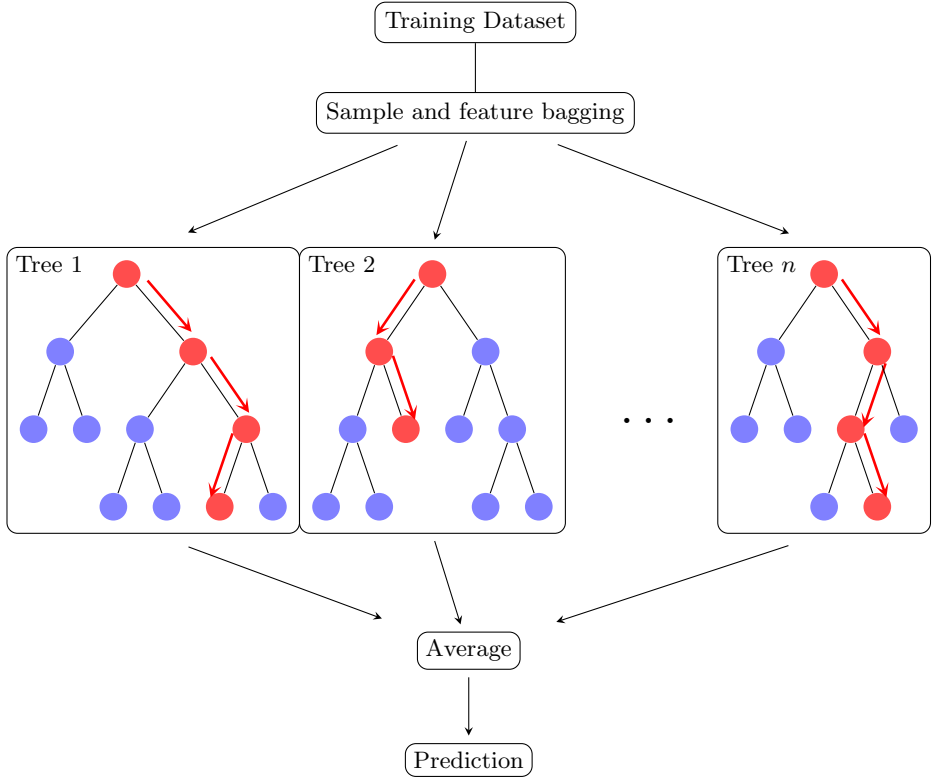
\begin{figure}
    \label{fig:rf_sketch}
    \centering
    \begin{forest}
    for tree={l sep=2em, s sep=1em, anchor=center, inner sep=0.4em, fill=blue!50, circle, where level=2{no edge}{}}
    [
    Training Dataset, node box
    [Sample and feature bagging, node box, alias=bagging, above=3em
    [,red!70,alias=a1[[,alias=a2][]][,red!70,edge label={node[above=1ex,red arrow]{}}[[][]][,red!70,edge label={node[above=1ex,red arrow]{}}[,red!70,edge label={node[below=1ex,red arrow]{}}][,alias=a3]]]]
    [,red!70,alias=b1[,red!70,edge label={node[below=1ex,red arrow]{}}[[,alias=b2][]][,red!70,edge label={node[above=1ex,red arrow]{}}]][[][[][,alias=b3]]]]
    [~~~$\dots$~,scale=2,no edge,fill=none,yshift=-3em]
    [,red!70,alias=c1[[,alias=c2][]][,red!70,edge label={node[above=1ex,red arrow]{}}[,red!70,edge label={node[above=1ex,red arrow]{}}[,alias=c3][,red!70,edge label={node[above=1ex,red arrow]{}}]][,alias=c4]]]]
    ]
    \node[tree box, fit=(a1)(a2)(a3)] (t1) {};
    \node[tree box, fit=(b1)(b2)(b3)] (t2) {};
    \node[tree box, fit=(c1)(c2)(c3)(c4)] (tn) {};
    \node[below right=0.5em, inner sep=0pt] at (t1.north west) {Tree 1};
    \node[below right=0.5em, inner sep=0pt] at (t2.north west) {Tree 2};
    \node[below right=0.5em, inner sep=0pt] at (tn.north west) {Tree $n$};
    \path (t1.south west)--(tn.south east) node[midway,below=4em, node box] (mean) {Average};
    \node[below=3em of mean, node box] (pred) {Prediction};
    \draw[black arrow={5mm}{4mm}] (bagging) -- (t1.north);
    \draw[black arrow] (bagging) -- (t2.north);
    \draw[black arrow={5mm}{4mm}] (bagging) -- (tn.north);
    \draw[black arrow={5mm}{5mm}] (t1.south) -- (mean);
    \draw[black arrow] (t2.south) -- (mean);
    \draw[black arrow={5mm}{5mm}] (tn.south) -- (mean);
    \draw[black arrow] (mean) -- (pred);
    \end{forest}
    \abovecaptionskip=12pt 
    \caption{RF algorithm diagram adapted from \citet{janosh_tikz_2023}. The paths highlighted in red represent the series of decisions made by each tree in the forest to make the final prediction. The RF prediction is the average of the trees final decisions. Each node represents a logical operation on a single feature. For example, a node may be asking if the variable E300 $<$ 100 (cm$^{2}$ s sr MeV)$^{-1}$. If yes go to the next node on the right, otherwise go to the next node on the left. The graph and the logical operations are inferred from data during the training phase of the model.}
\end{figure}

The algorithm used to build the model is the Random Forest (RF).
The RF is an ensemble learning algorithm that 
operates by creating an ensemble of decision trees through bootstrapped sampling and random feature selection. 
A decision tree is a hierarchical structure used in both classification and regression tasks. The training process of a decision tree involves recursively splitting the data according with feature values to make predictions. These splits are determined by optimization criteria such as the Gini impurity for classification or the mean squared error (MSE) for regression. Each internal node of a decision tree represents a feature and a decision rule associated with the feature, while each leaf node represents a prediction. For example, given a feature vector $\mathbf{X}$, the root node may be asking the value of the feature $i$-th feature, $x_i$. The algorithm reads the value $x_i$ and compare it to the threshold value $\epsilon$. If $x_i > \epsilon$, then the algorithm decides to go through the branch 1 composed by additional nodes checking the other features otherwise it goes through the branch 2 composed by some other nodes.  The hierarchical structure of each decision tree (i.e. the value of the thresholds $\epsilon_i$ and the nodes) is automatically created during the learning phase.
However, a decision tree alone is susceptible to overfitting when the tree becomes too complex and closely fits the training data. The RF algorithm has been introduced with the aim of overcoming this problem. The idea of the RF algorithm is to create an ensemble of independent decision trees, each one making its own prediction (after the learning phase). Then the final prediction is the arithmetic average of decision trees outcomes (see Figure \ref{fig:rf_sketch} for a sketch).

The RF algorithm can be effectively adapted to timeseries forecasting, by
redefining the problem as a regression task. To achieve this, we create a modified dataset where the target variable $y$ is paired with lagged versions of each single feature $x$, shifted forward in time by a fixed amount $t' = n$. In this setup, $y_t$ is compared to $x_{t-n}$, $y_{t+1}$ is compared to $x_{t-n+1}$, and so forth. In essence, we establish a relationship between the current target variable $y_t$ and the historical values of each feature $x$, allowing the model to make predictions of $y$'s values $n$ time steps into the future.
By training the RF on this modified dataset, the model learns the dependencies between the past values of each feature $x$ and future values of $y$ and automatically builds the hierarchical structure of each decision tree. Formally the model is expressed by a function $f$ mapping the total feature vector $\mathbf{X}_{t, t_b}$ into the target $y_t$, i.e., 
\begin{equation}\label{eq:formal_model}
   y_{t+t_f} = f(\mathbf{X}_{t, t_b}; \mathbf{W}),
\end{equation}
where $t_f$ is the lookforward and $t_b$ is the lookback.
$\mathbf{W}$ is a set of parameters that define the structure of the trees (nodes, thresholds and the associated decision rules) and are learnt from the training dataset. Note that, for a fixed time $t$ the features are represented by the vector $\mathbf{X}$, which incorporates the lags of each physical observable (i.e., past values of the observables are flattened in a $N$-dimensional row vector, where $N$ is the total number of the features). Practically, in this way, we are able to define the lookback $t_b$ for the RF. Therefore, for example, suppose to have a single feature $x$. Then the features vector would be $\mathbf{X}_{t, t_b}=(x_t, x_{t-t_1}, ..., x_{t-t_b})$, where $t_i$ denotes the i-th lag of the observable $x$, i.e., $t_i \in \left[t_1, t_b\right]$. The whole vector $\mathbf{X}_{t, t_b}$ is associated to the single output $y_{t+t_f}$, i.e., 
\begin{equation}
    \mathbf{X}_{t, t_b} \mapsto y_{t+t_f}.
\end{equation}
In principle, the lags $t_i$ need not to be equally spaced.

Finally, note also that the function defined in Eq. \eqref{eq:formal_model} implicitly depends on a set of hyperparameters which control for example the number of trees in the forest or the loss function used for optimizing the weights $\mathbf{W}$.
In the following we are going to consider the hyperparameters reported in Table \ref{tab:tab_rf_pars} and a single-output model. 

\section{Results}

To asses the model performance, the dataset has been split in two parts. The first part used for the learning phase (training dataset) and the second one used for testing only (test dataset). Historical electron and proton flux timeseries have been considered in the analysis, i.e., without separating transient events (SPEs) from the background, thus enabling the model to accurately learn the differences between them. The dataset has been used with full cadence sampling of 1 minute.
The training dataset ranges from 1996 January 1 to 2012 February 12, whereas the test dataset ranges from 2012 February 13 to 2017 September 2. Data from 2017 October 4 onwards are not included because of instrument reconfiguration (see \url{http://www2.physik.uni-kiel.de/SOHO/phpeph/EPHIN.htm}).

\begin{figure}[!h]
   	\centering
   	\includegraphics[width=\columnwidth]{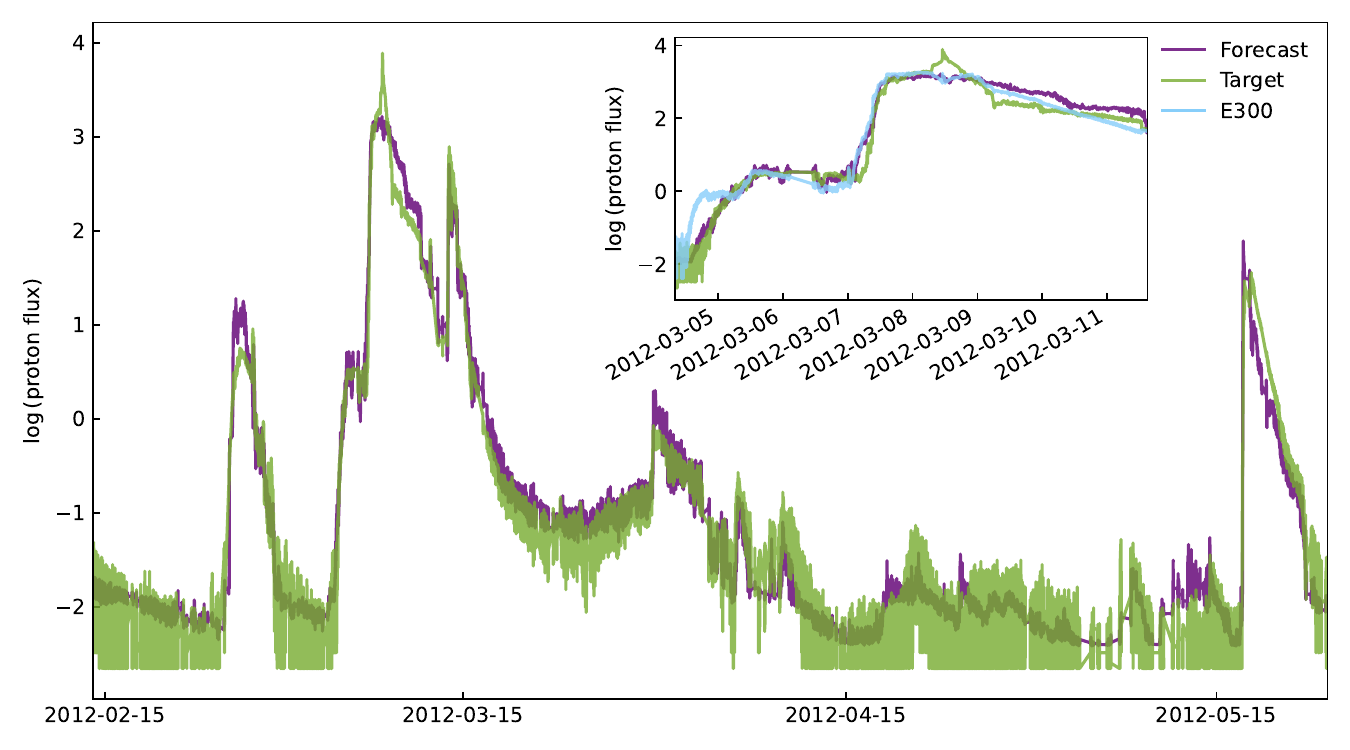}
   	\caption{A 3-months sample dataset used for testing/visualization purposes. Green line represents the target of the model (true values), while blue line represents the output of the model (forecasting). Both are representative of the logarithm of the proton flux integrated from 7.8 MeV and 53 MeV energy channels (P8, P25, P41; see Table \ref{Tab:tab1}). The proton flux is given in pfu, while the logarithm is adimensional. The inset shows a closer look around the ESP event of 2012-03-08, which is not well reproduced by the model. This is due to the fact that the electrons do not exhibit the ESP signature (see the light blue line, which shows the track of the E300 electron differential channel).}
   	\label{fig:fig_3}
\end{figure}

Figure \ref{fig:fig_3} shows a sample from the test dataset. The inset panel of the same Figure shows a closer view of the event of 2012 March 7. 
All the SPEs observed in the time intervals depicted in Figure \ref{fig:fig_3} are precisely reproduced from the model in terms of both proton flux value and timing, i.e., event onsets and flux values.
By inspecting the SPE of March 2012 (inset of Figure \ref{fig:fig_3}), the overall trend of the proton flux is nicely reproduced by the model, which provides also realistic fluctuations overlaid on a global trend of the event. Hence, the features selected for the model provide extensive information about energetic proton flux characteristics. However, such features are mainly derived from electron flux measurements and the proton derivative, thus being informative of the location/magnetic connectivity of the probe with the Sun and of the particle acceleration processes at the source. Signatures of the proton local energization associated with shock propagation, i.e., the so called energetic storm particles (ESP), are then not expected in the predicted timeseries. As a confirmation of this, the EPHIN in situ measurements of the March 2012 event reported in the inset panel of Figure \ref{fig:fig_3} exhibit the typical peak of an ESP event, which, conversely, is not present in the predicted proton flux.

To better asses the quality of the predictions we refer to: i) the root mean squared error (RMSE), which is in the same units as the target; ii) the coefficient of determination $R^2$; iii) the cross correlation between the target timeseries and the predicted one. The i) and ii) provide the performance metrics which quantify the information explained by the model, i.e., how well the details of the target time series (e.g. trend and fluctuations) are reproduced by the predicted one.
The RMSE is defined as the mean distance between the target $y_{t_k+n}$ and the predicted value $\hat{y}_{t_k+n}$
\begin{equation}\label{eq:rmse}
   \text{RMSE} =  \sqrt{\frac{\sum_{k=1}^N (y_{t_k+n}- \hat{y}_{t_k+n})^2}{N}},
\end{equation}
and provides a measure of the mean amplitude of the fluctuations around the target $y_{t_k+n}$. 

The coefficient of determination $R^2$ is in essence the explained variance of the predictions and is defined as
\begin{equation}
    R^2 = 1 - \frac{\sum_{k=1}^N(y_{t_k+n}- \hat{y}_{t_k+n})^2}{\sum_{k=1}^N (y_{t_k+n}- \bar{y})^2},
\end{equation}
where $\bar{y}$ is the mean of the target. The value of $R^2$ computed on a dataset (not used for fitting the model) ranges from $-\infty$ to 1, where in particular $R^2=1$ means perfect predictions. 

Finally, we computed the iii) cross correlation for measuring the mean time shift between the prediction and the target timeseries and is defined by
\begin{equation}\label{eq:cross_corr}
    C(\tau) \propto \langle y(t+\tau)\hat{y}(t) \rangle,
\end{equation}
provided that $y$ and $\hat{y}$ timeseries have been subtracted to their mean values. We used the cross correlation function to measure the warning time statistically. In particular, we defined the warning time $\Delta t$ as a function of the lag $\bar{\tau}$ that maximizes the cross correlation function, i.e., $\Delta t = t_f - \bar{\tau}$ is given by the lookforward of the model (which is $t_f = 1$ hour), minus the lag $\bar{\tau}$. Thus, by definition, the warning time $\Delta t$ may be $\Delta t \leq 1$ hr.

\begin{figure}[!h]
   	\centering
    \includegraphics[width=\columnwidth]{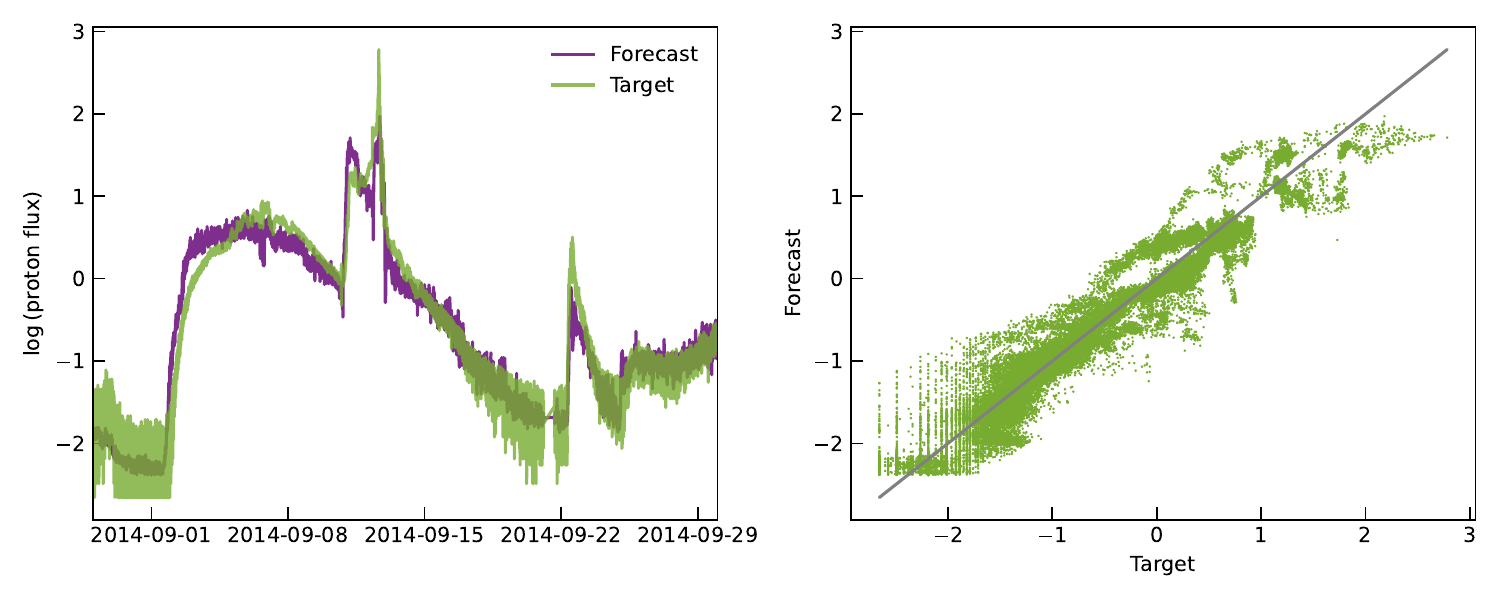}
    \includegraphics[width=\columnwidth]{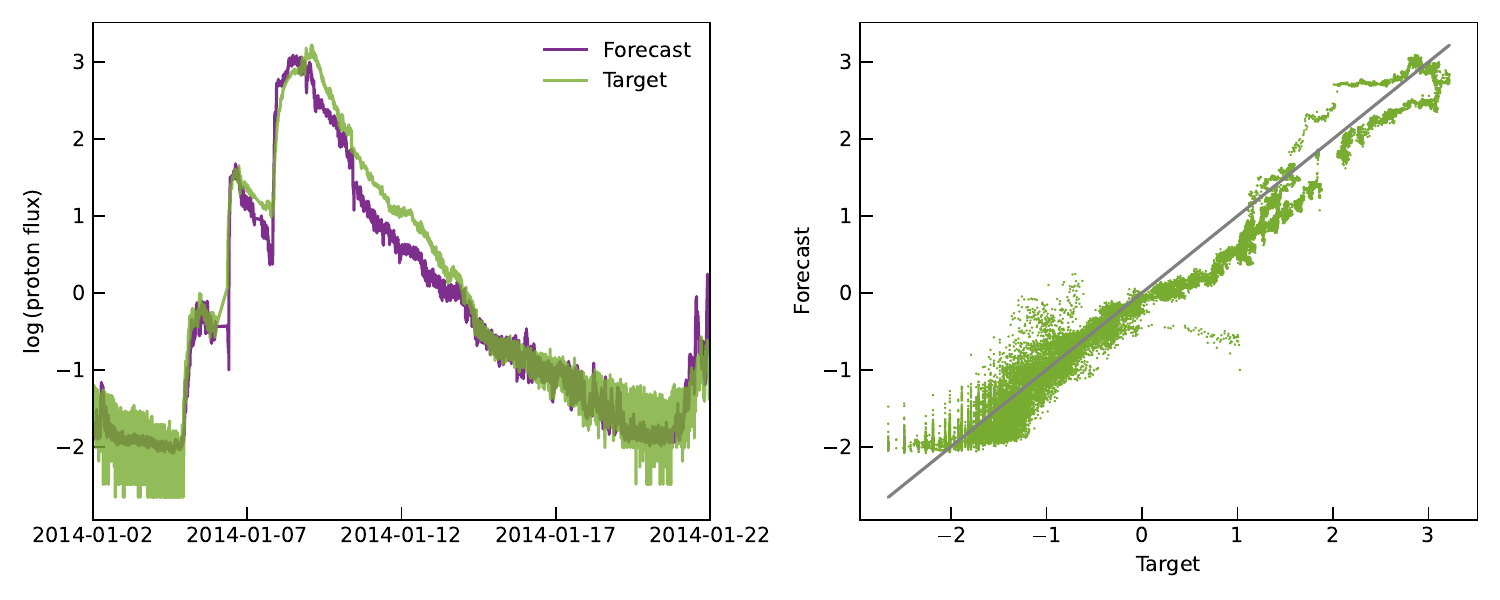}
    \includegraphics[width=\columnwidth]{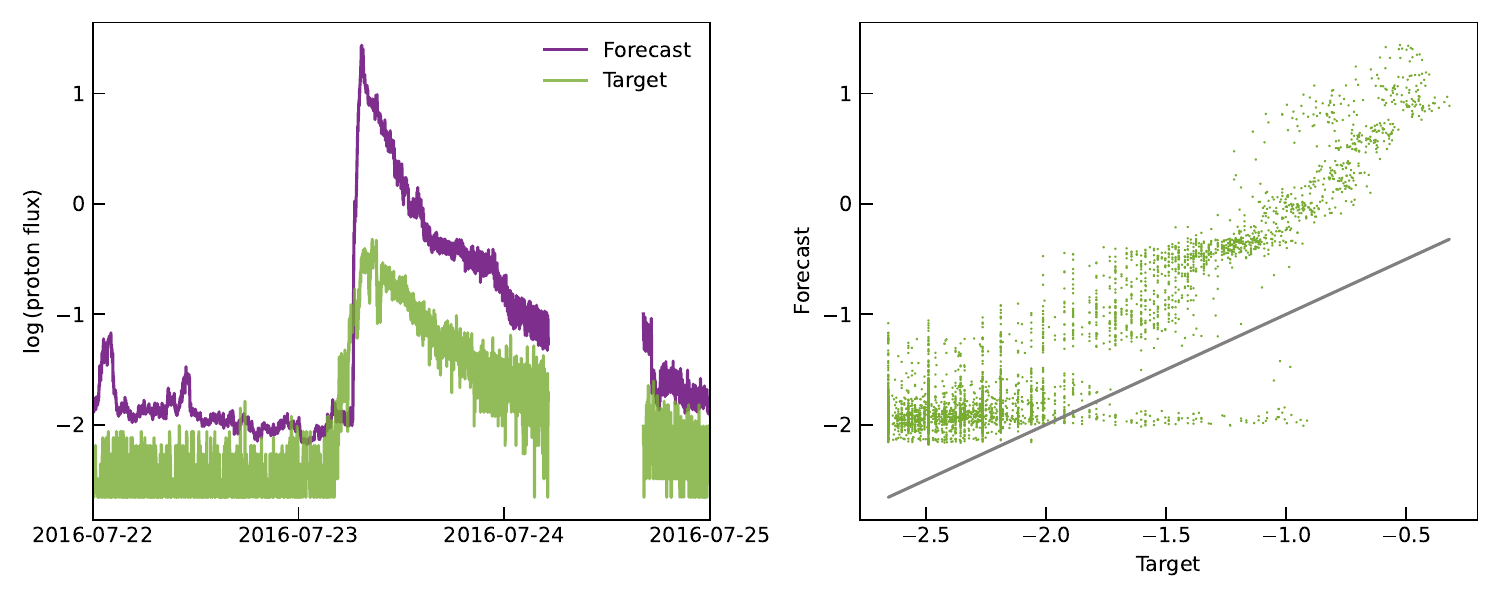}
    \caption{(Left panels) Comparison between the predicted time series and the target one for three samples: 23, 18 and 31 (from top to bottom). (Right panel) correlation plot between the predicted and the target points. The grey line represents the bisector (not the linear regression line). Note that the sample 31 is the worst possible example (see Table \ref{Tab:tab_summary}). However, even though the model overestimates the flux, the trend is reproduced quite well during the enhancement of 2016-07-23.}
   	\label{fig:fig_5}
\end{figure}

To test the model we selected a set of testing samples. Each sample contains the signature of a proton flux enhancement, not necessarily corresponding to the overcoming of the 10 pfu threshold (defined for GOES proton flux at $>10$ MeV). Furthermore, the samples used for testing have been excluded from the training dataset, in order to avoid any bias in the estimation of performance metrics.
The results are summarized in Table \ref{Tab:tab_summary} where, from left to right, we report: the start date, the end date, the coefficient of determination $R^2$, the RMSE, the peak flux reached by the target time series in the selected time interval, the peak reached by the predicted time series in the same time interval, and the statistical warning time based on the estimation of the cross correlation function. 
Generally, the model shows a quite good performance in terms of $R^2$, with the exception of sample 31, for which $R^2 = -0.49$ is negative. However, the bottom panel of Figure \ref{fig:fig_5}, which compares the model's prediction with the target for sample 31, shows that even though the flux is overestimated, the trend is quite well reproduced, especially during the enhancement of the proton flux observed on 2016-07-23. In terms of the RMSE, which quantifies the fluctuations of the prediction with respect to the target, the model has better performance for higher values of the proton flux (see e.g. samples 12 and 34). This may be due to the effect of the log-transformation. Indeed, as can be also seen from Figure \ref{fig:fig_3}, the fluctuations of the target are generally more pronounced for lower values. 

The warning time $\Delta t$ is predominantly greater than zero, meaning that the model forecast the enhancement of the proton flux in advance. Almost for all the testing samples the warning time is equal to the lookforward of the model (1 hr), i.e., predicted and target timeseries are synchronized on average. The only exception is sample 24 which is not timely predicted. 

Figure \ref{fig:fig_5} shows the results in terms of the timeseries of three test samples (e.g. from top to bottom, 23, 18 and 31). The left, top and middle panels show two testing samples for which the model is quite good in terms of estimation of the flux and timing of both the onset and the peak. The right top and middle panels of each row also show the correlation plot between the prediction and the target. As can be seen, the points are almost distributed around the bisector, meaning that the target and the prediction are comparable. In the bottom panel, however, we show the worse sample in terms of model performance (sample 31). Even though the flux is here overestimated and the points are not distributed along the bisector in the correlation plot (right bottom panel of Figure \ref{fig:fig_5}), the enhancement is still predicted by the model as can be seen also from the fact that the points are linearly related.

    \begin{table}[!h]
        \centering
        \begin{tabular}{c|c|c|c|c|c|c|c}
            Sample N. & Start date & End date & $R^2$ & RMSE & target peak & pred. peak & Warning time [min]\\
            \hline
            1 & 2012-02-24 & 2012-03-01 & 0.92 & 0.33 & 0.96 & 1.28 & 60\\
2 & 2012-03-04 & 2012-03-22 & 0.97 & 0.26 & 3.89 & 3.22 & 60\\
3 & 2012-05-16 & 2012-05-24 & 0.93 & 0.34 & 1.55 & 1.89 & 60\\
4 & 2012-05-26 & 2012-06-01 & 0.94 & 0.25 & 1.21 & 0.80 & 60\\
5 & 2012-06-14 & 2012-06-21 & 0.71 & 0.56 & 1.77 & 0.34 & 60\\
6 & 2012-07-05 & 2012-08-05 & 0.81 & 0.37 & 2.25 & 1.69 & 20\\
7 & 2012-08-27 & 2012-09-18 & 0.89 & 0.41 & 1.97 & 1.58 & 60\\
8 & 2012-09-17 & 2012-10-13 & 0.88 & 0.36 & 1.25 & 1.35 & 60\\
9 & 2013-02-21 & 2013-03-02 & 0.69 & 0.46 & 0.13 & -0.13 & 60\\
10 & 2013-03-05 & 2013-04-01 & 0.93 & 0.28 & 1.54 & 1.05 & 60\\
11 & 2013-04-08 & 2013-05-01 & 0.91 & 0.34 & 2.00 & 1.58 & 60\\
12 & 2013-05-09 & 2013-06-05 & 0.94 & 0.39 & 3.19 & 3.14 & 60\\
13 & 2013-06-21 & 2013-07-03 & 0.93 & 0.32 & 1.27 & 0.89 & 60\\
14 & 2013-08-15 & 2013-09-08 & 0.87 & 0.26 & 0.35 & 0.24 & 60\\
15 & 2013-09-28 & 2013-10-10 & 0.96 & 0.31 & 2.55 & 2.23 & 60\\
16 & 2013-10-11 & 2013-11-25 & 0.83 & 0.34 & 0.72 & 1.40 & 60\\
17 & 2013-12-27 & 2013-12-31 & 0.48 & 0.44 & 1.43 & 0.52 & 60\\
18 & 2014-01-02 & 2014-01-22 & 0.96 & 0.30 & 3.22 & 3.09 & 60\\
19 & 2014-02-16 & 2014-02-24 & 0.80 & 0.40 & 1.43 & 1.53 & 30\\
20 & 2014-02-24 & 2014-03-21 & 0.97 & 0.26 & 2.18 & 2.24 & 60\\
21 & 2014-03-28 & 2014-04-02 & 0.84 & 0.28 & 0.45 & 1.06 & 59\\
22 & 2014-04-16 & 2014-04-30 & 0.90 & 0.43 & 2.17 & 1.66 & 60\\
23 & 2014-08-29 & 2014-09-30 & 0.94 & 0.26 & 2.78 & 1.97 & 60\\
24 & 2014-10-31 & 2014-11-08 & 0.90 & 0.36 & 1.28 & 0.77 & 0\\
25 & 2015-06-17 & 2015-07-08 & 0.83 & 0.58 & 3.45 & 2.40 & 60\\
26 & 2015-09-18 & 2015-09-26 & 0.89 & 0.25 & 0.32 & 0.66 & 60\\
27 & 2015-10-26 & 2015-11-04 & 0.91 & 0.32 & 1.00 & 1.56 & 60\\
28 & 2015-12-27 & 2015-12-31 & 0.68 & 0.60 & 0.76 & 0.17 & 60\\
29 & 2016-01-01 & 2016-01-07 & 0.78 & 0.56 & 1.80 & 0.70 & 27\\
30 & 2016-05-15 & 2016-06-19 & 0.56 & 0.32 & 0.41 & 0.07 & 60\\
31 & 2016-07-22 & 2016-07-25 & -0.49 & 0.73 & -0.32 & 1.49 & 41\\
32 & 2017-07-13 & 2017-07-19 & 0.91 & 0.41 & 1.42 & 1.09 & 60\\
33 & 2017-07-22 & 2017-08-06 & 0.88 & 0.27 & 0.15 & 0.12 & 60\\
34 & 2017-09-02 & 2017-09-30 & 0.90 & 0.48 & 3.30 & 3.17 & 32\\
35 & 2021-10-28$^{*}$ & 2021-11-05$^{*}$ & 0.84 & 0.45 & 0.87 & 1.08 & 58\\
36 & 2022-03-28$^{*}$ & 2022-04-07$^{*}$ & 0.82 & 0.46 & 0.86 & 1.50 & 48\\
            \hline
        \end{tabular}
        \caption{Error metrics for sample events compared with the peak flux. From left to right: the start date, the end date, the coefficient of determination $R^2$, the RMSE, the target peak, the predicted peak and the warning time. Note that the target peak, the predicted peak and the RMSE are referred to the logarithm of the proton flux. Thus, given for example the predicted peak $x$, the proton flux in pfu is $10^x$. The warning time $\Delta t$ is computed as the difference between the lookforward and the cross correlation lag, i.e., $\Delta t = t_f - \bar{\tau}$. We also included two events after 2017 October 4 (denoted by asterisks). The results for these two samples have to be taken with care.}
        \label{Tab:tab_summary}
    \end{table}

\section{Discussion and conclusion}
In this work we have presented a pipeline for predicting the energetic proton flux by using a machine learning model based on the RF algorithm. We considered data from the SOHO/EPHIN telescope for training and testing the model. The reason why we used the SOHO/EPHIN data for demonstrating the effectiveness of our approach is that differential channels for both protons and electrons are embedded in the same dataset, so there are no possible effects due to spacecraft placed in different environments. The pipeline for generating the most informative dataset resulted in a set of 27 features, all extracted from the electron flux channels and derivative of the integrated proton flux channel, both with different time lags. 
The model average performance is reported in Table \ref{Tab:tab_mean}. All the metrics reported suggest that the model has a remarkable performance in predicting the energetic proton flux up to one hour ahead by exploiting information derived from the electron differential channels. Nevertheless, we remark that these are the average characteristics of the model.
There may be some events anticipating or delaying the characteristics of the time profile, making a non-timely prediction of the onset or of the peak (see Table \ref{Tab:tab_summary}). This may be due to the fact that the delay between protons and electrons (onset or peak) changes dynamically event per event, depending on several factors, such as: the magnetic connectivity with the source of particles, the different release time and/or sources of electrons and protons, and their propagation in the interplanetary space \citep{Kollhoff2021,Dresing2023}. According to \citet{posner2007up} the distribution of the delay between protons and electrons onsets range from 20-30 minutes up to 1 hour. 
However, in this work, we took into account the dependence of the delay on the magnetic connectivity by employing the time derivative of both the electron and proton differential channels, which in turn contain the information about the rise parameter. The resulting mean warning time (averaged over the samples reported in Table \ref{Tab:tab_summary}) is 54 minutes.

At its current state, the model produces a single output $y_{t+t_f}$. 
In principle, a multi-output regressor can represent a natural evolution of the RF-based model presented in this work, providing a vector $(y_{t+1}, y_{t+2}, ..., y_{t+t_f})$ which contains the prediction of the time profile in the interval $[t+1,t+t_f]$, where $t_f$ indicates the size of the forecasting window (i.e. the lookforward). This perspective could be particularly helpful for having an idea of SPEs duration and generally more accurate information on the timings.

From a physical perspective, the predictability of the proton flux from the electron flux is plausible, since protons and electrons propagation is along the same interplanetary magnetic field configuration (a spiral during non disturbed periods), given that they are accelerated during the same solar eruption, with electrons arriving earlier than protons at the observer location due to the mass difference. Hence, electrons may carry information about the value of proton flux and the magnetic connectivity of the observer with the source \citep{posner2007up}.
On the contrary, characteristics of the proton flux associated with local proton/ion energization processes, such as ESPs usually observed during interplanetary shock passages, are not present in the electron flux channels and then are not reproduced by the model.
\begin{table}[!h]
    \centering
    \begin{tabular}{c|c|c}
        $R^2$ & RMSE & Warning time [min]\\
        \hline
        0.82 & 0.38 & 54\\
        \hline
    \end{tabular}
    \caption{Error metrics for the model averaged over the test samples reported in Table \ref{Tab:tab_mean}. From left to right: coefficient of determination $R^2$, root mean squared error and the warning time. Note that the RMSE is referred to the logarithm of the proton flux.}
    \label{Tab:tab_mean}
\end{table}

In this work we used specific differential channels for both electrons and protons, therefore our model is very specialized to predict the protons in the EPHIN integral channel. However, these results highlight that a very simple algorithm such as the RF is able to capture the variability of the solar proton flux by extracting basic features (mean, standard deviation and derivative) from the electron channels. Thus we may speculate that this approach can possibly be valid also for other instruments offering similar observations. In those cases, possible differences in the performance would depend mainly on the quality of the available data.
This makes our approach very promising in view of space missions carrying on board particle detectors that allows for electron and proton flux measurements, with the option of providing a dedicated SPEs prediction system to be directly implemented on board. In principle, the model can be coupled with a decision rule to issue an alert. For example the most naive approach would be to issue an alert if the predicted value overcome a threshold $\epsilon$. Another approach could be to issue an alert when the predicted time series presents an anomalous value. More sophisticated approach are also feasible in terms of issuing alerts, like anomaly detection algorithm running directly on the predicted timeseries. However, the advantage of predicting the future values of the proton flux instead of giving the classification as output is that the decision rule can be adapted to any situation and also allows more detailed analysis of the trend.

Therefore, this work is particularly relevant for the implementation of Machine Learning algorithms for the next generation applications on board space missions, which is a critical need for the future of space exploration. For instance, the development of AI-based algorithms for the automation of on board operations, including SPE forecasting, is a main goal of the ASAP (Automatics in SpAce exPloration; \url{https://asap-space.eu}) project, funded by the Horizon Europe Framework Programme.
Our system can be also easily adapted aboard the HENON (HEliospheric pioNeer for sOlar and interplanetary
threats defeNce) pathfinder mission dedidated to Space Weather operations and science. HENON is currently in the detailed design Phase C1, foreseeing a cubesat on the Distant Retrograde Orbit in deep space well
upstream of L1 around 0.1 AU toward the Sun. Being equipped with a radiation monitor, a Faraday cup and a magnetometer, HENON allows, in principle, the early computation and near real time transmission to Earth of alerts for the Space Weather events, including SPEs 
\citep{Laurenza2023,provinciali2024henon}.
In addition, an onboard SPE forecasting model is essential to address the issue of astronaut radiation safety and to limit human exposure to high‐dose rates for space exploration missions to the Moon and Mars. 
Possible forecasting systems relying on remote sensing observations could face the problem at Mars distance that for up to half of the relevant SPE events, the nominal magnetic ﬁeld line connection to the Sun is at ~W90, implying that the full flare signature may not be visible and obstructed from view. Forecasting systems based only on near-relativistic observations, such as a two-element REleASE warning system \citep{posner2020warning} and the model presented here, can provide a valid solution as they are not affected by this issue.

\section*{Acknowledgements \label{sec:ack}}
This study has been performed in the framework of the HEliospheric pioNeer for sOlar and interplanetary
threats defeNce (HENON) mission Phase A/B  and C1. HENON is part of the Italian Space Agency (ASI) program Alcor and is being developed under the European Space Agency General Support Technology Programme (ESA-GSTP) through the support of the national delegations of Italy (ASI), UK, Finland and Czech Republic. MS acknowledges the support from the ASI-SERENA contract no. 2018-8-HH.O ‘‘Partecipazione scientifica alla missione BEPICOLOMBO SERENA Fase E" and the ESA contract (RFP/NC/ IPL- PSS/JD/258.2016) Expert Support to SERENA Science Operations. MS, ML, SB and MFM acknowledge the SOHO/EPHIN team for providing the data used in this work.

\bibliography{stumpoetal}{}
\bibliographystyle{aasjournal}

\end{document}